\documentclass[reqno, a4paper]{amsart}
\usepackage{amsmath}
\usepackage{amssymb}
\usepackage{amsthm}

\usepackage[scale=0.8]{geometry}

\usepackage{natbib}
\usepackage{bibentry} 

\usepackage[english]{babel}
\usepackage[utf8]{inputenc}

\usepackage{subfig}
\usepackage{graphicx}

\usepackage[unicode,breaklinks]{hyperref}
\usepackage[active]{srcltx} 

\usepackage{mathbbol}
\usepackage{bm} 
\usepackage{stmaryrd}
\usepackage{MnSymbol} 
\usepackage{times}

\usepackage{units}
\usepackage{tensor}
\usepackage{accents}

\usepackage{enumitem}

\usepackage{lineno}

\newcommand{\exponential}[1]{\ensuremath{{\mathrm e}^{#1}}}

\newcommand{\bydefinition}{\mathrm{def}}

\newcommand{\diff}{\mathrm{d}}

\makeatletter
\@ifpackageloaded{bm}%
{%
}{%
\relax
}
\makeatother

\newcommand{\inverse}[1]{#1^{-1}}

\makeatletter
\@ifpackageloaded{bm}%
{%
}{%

}

\@ifpackageloaded{bm}%
{%
 
}{%

}

\makeatother

\newcommand{\diracdelta}{\delta}
\newcommand{\Heaviside}{H}

\newcommand{\R}{\ensuremath{{\mathbb R}}}

\newcommand{\N}{\ensuremath{{\mathbb N}}}

\newcommand{\dd}[2]{\ensuremath{\frac{\diff {#1}}{\diff {#2}}}}

\newcommand{\ddd}[2]{\ensuremath{\frac{\diff^2 {#1}}{\diff {#2}^2}}}

\makeatletter
\@ifpackageloaded{tensor}
{

}{%

}
\makeatother

\makeatletter
\@ifpackageloaded{tensor}
{

}{%

}
\makeatother

\makeatletter
\@ifundefined{volume}{%
}%
{%
}
\makeatother

\makeatletter
\@ifpackageloaded{MnSymbol} 
{
 
}{%
 
}
\makeatother

\newtheorem{theorem}{Theorem} 

\newtheorem{lemma}[theorem]{Lemma}

\theoremstyle{proposition}

\newcommand{\genfn}[1]{{\boldsymbol {#1}}}

\newcommand{\equilibrium}{\mathrm{eq}}
\newcommand{\final}{\mathrm{jp+}}
\let\cite\citet
\numberwithin{equation}{section}

\title[Step load and deformation in nonlinear mechanical systems]{Colombeau algebra as a mathematical tool for investigating step load and step deformation of systems of nonlinear springs and dashpots}

\author{V\'{\i}t Pr\r{u}\v{s}a}
\date{\today}
\address{%
Faculty of Mathematics and Physics\\
Charles University in Prague\\
Sokolovsk\'a 83\\
Praha 8 -- Karl\'{\i}n\\
CZ 186\;75\\
Czech Republic
}
\email{prusv@karlin.mff.cuni.cz}

\author{Martin \v{R}eho\v{r}} 
\address{%
Faculty of Mathematics and Physics\\
Charles University in Prague\\
Sokolovsk\'a 83, Praha 8 -- Karl\'{\i}n\\
CZ 186\;75, Czech Republic
}
\email{rehor@karlin.mff.cuni.cz}

\author{Karel T\r{u}ma}
\address{%
Institute for Fundamental Technological Research\\
Polish Academy of Sciences\\
Adolfa Pawi\'nskiego 5B\\
Warszawa\\
PL 02-106\\
Poland
}
\email{ktuma@ippt.pan.pl}

\thanks{V\'{\i}t Pr\r{u}\v{s}a was supported by the project LL1202 in the programme ERC-CZ funded by the Ministry of Education, Youth and Sports of the Czech Republic.}

\begin{document}

\begin{abstract}
The response of mechanical systems composed of springs and dashpots to a step input is of eminent interest in the applications. If the system is formed by linear elements, then its response is governed by a system of linear ordinary differential equations, and the mathematical method of choice for the analysis of the response of such systems is the classical theory of distributions. However, if the system contains nonlinear elements, then the classical theory of distributions is of no use, since it is strictly limited to the linear setting. Consequently, a question arises whether it is even possible or reasonable to study  the response of nonlinear systems to step inputs. The answer is positive. A mathematical theory that can handle the challenge is the so-called Colombeau algebra. Building on the abstract result by (Pr\r{u}\v{s}a \& Rajagopal 2016, Int. J. Non-Linear Mech) we show how to use the theory in the analysis of response of a simple nonlinear mass--spring--dashpot system.


\end{abstract}

\maketitle



\section{Introduction}
\label{sec:introduction}
The behaviour of systems governed by ordinary differential equations is of interest in many branches of mechanics. A~prominent example of mechanical systems governed by ordinary differential equations are systems composed of springs and dashpots, which are of interest either by themselves or they can serve as reduced models of more complex systems. 

In the applications one frequently needs to determine the response of spring-dashpot systems to a step input, which can be either a step loading or a step deformation. Since the behaviour of these systems is described in terms of ordinary \emph{differential} equations, and the step input \emph{lacks differentiability}, one immediately sees that the study of step inputs requires a proper generalisation of the standard notion of the derivative of a function.

If the system of interest is formed by linear elements, then the governing ordinary differential equations are \emph{linear}. In such a case the classical theory of distributions, see~\cite{schwartz.l:theorie}, provides a suitable tool for extending the concept of the derivative even for discontinuous objects. However, if the elements of the spring--dashpot system are nonlinear, then the governing ordinary differential equations are also \emph{nonlinear}, and \emph{the classical theory of distributions is of no use} since it is essentially limited to the linear setting. In particular, the classical theory of distributions does not provide one a suitable definition of the product of two distributions (generalised functions), and it even seems that a theory extending the classical theory of distributions to a nonlinear setting can not exist at all, see for example the celebrated ``impossibility result'' by~\cite{schwartz.l:sur}.  

Moreover, if one is willing to ignore mathematical rigour, and extend the classical theory to a nonlinear setting by appealing to common sense and the standard calculus rules, then one immediately ends up with paradoxical results. For example, a \emph{naive} calculation based on the \emph{apparently obvious} equality $\Heaviside^m = \Heaviside^n$, where $\Heaviside$ is the Heaviside \emph{step} function 
\begin{equation}
  \label{eq:1}
  \Heaviside(t)
  =_{\bydefinition}
  \begin{cases}
    0, & t < 0, \\
    1, & t \geq 0,
  \end{cases}
\end{equation}
and $n, m \in \N$, $m\not=m$, would imply the following. The differentiation of $\Heaviside^m = \Heaviside^n$ would yield
$
  m \Heaviside^{m-1} \dd{\Heaviside}{t} =n \Heaviside^{n-1} \dd{\Heaviside}{t},
$
and consequently one would get
\begin{equation}
  \label{eq:3}
  m \dd{\Heaviside}{t} = n \dd{\Heaviside}{t},
\end{equation}
which is absurd. Consequently, if one can not handle nonlinear operations with Heaviside function that is the simplest possible function describing a step input, then it seems that the case is lost, and that one can not handle the step input in the nonlinear setting at all. 

Fortunately, the contrary is true. There exists a generalisation of the theory of distributions---the so-called \emph{Colombeau algebra}---that is suitable for the nonlinear setting, see~\cite{colombeau.j:new,colombeau.j:multiplication} and \cite{rosinger.ee:generalized,rosinger.ee:nonlinear}. 

The classical ``impossibility result'' by~\cite{schwartz.l:sur} is still valid and no contradiction arises. The construction of Colombeau algebra is based on the weakening of one of the requirements imposed by~\cite{schwartz.l:sur} to the hypothetical nonlinear theory of distributions. The requirement that has been found too restrictive is the requirement on compatibility of the classical multiplication and the multiplication in the hypothetical nonlinear theory. \cite{schwartz.l:sur} required the multiplication in the hypothetical nonlinear theory to coincide with the classical multiplication provided that one considers \emph{continuous} functions. If this requirement is weakened to the compatibility of the multiplication only for \emph{smooth} functions, then a nonlinear theory of distributions can be introduced. Consequently, the price to pay to overcome the limitations of the ``impossibility result'' is a complex structure of Colombeau algebra. 

On the other hand, the complexity of Colombeau algebra is a natural consequence of the complexity of the problem. It would be futile to expect an existence of an extremely simple theory that can simultaneously handle discontinuity, differentiation and nonlinearity. More importantly, the complexity of Colombeau algebra does not prevent one from using it in the applications. 

Indeed, using the calculus rules in Colombeau algebra \cite{prusa.v.rajagopal.kr:on} have been able to solve the problem of the response $\sigma(t)$ of systems governed by general nonlinear differential equations of the type
\begin{subequations}
  \label{eq:4}
  \begin{align}
    \label{eq:5}
    a(\varepsilon, \sigma) \sigma + \dd{\sigma}{t} &= b(\varepsilon, \sigma) \varepsilon + c(\varepsilon, \sigma) \dd{\varepsilon}{t}, \\
    \label{eq:6}
    \sigma + a(\sigma, \varepsilon) \dd{\sigma}{t} + b \ddd{\sigma}{t}  &=2 c(\sigma, \varepsilon) \dd{\varepsilon}{t} + 2 d \ddd{\varepsilon}{t},
  \end{align}
\end{subequations}
to the step input $\varepsilon(t)=_{\bydefinition}\tilde{\varepsilon}(t) \Heaviside(t)$, where $\tilde{\varepsilon}$ is a given smooth function. (Here $a(\varepsilon, \sigma)$, $b(\varepsilon, \sigma)$ and $c(\sigma, \varepsilon)$ are given smooth functions, and $b$ and $d$ are constants.) In particular, \cite{prusa.v.rajagopal.kr:on} have shown that the response to the step input is the step response of the form $\sigma(t) = \tilde{\sigma}(t) \Heaviside(t)$ where $\tilde{\sigma}(t)$ is a well specified function. As one might expect, the governing equations~\eqref{eq:4} remain valid \emph{in the classical sense} apart from the point of jump discontinuity. (The pair $\tilde{\sigma}$ and $\tilde{\varepsilon}$ solves~\eqref{eq:5} or~\eqref{eq:6} respectively for $t>0$.) The difficult part of the problem is the specification of the initial conditions---or the jump conditions---at $t=0+$ that are necessary for ``restarting'' the classical solution after the step change in the input. This issue has been addressed by~\cite{prusa.v.rajagopal.kr:on}. 

The value of the result by~\cite{prusa.v.rajagopal.kr:on} lies in the fact that the initial conditions have been found to be  fully determined by the governing equation~\eqref{eq:5} or~\eqref{eq:6} respectively, provided that the governing equations are interpreted in the context of Colombeau algebra. \emph{There is no need to supply the initial conditions by appealing to some external piece of information, everything is encoded in the governing equations themselves.}

\begin{figure}[h]
  \centering
  \subfloat[\label{fig:nonlinear-solid} Nonlinear spring--dashpot system.]{\includegraphics[height=0.2\textwidth]{figures-explanation/nonlinear-solid-explicit}}
  \qquad
  \subfloat[\label{fig:nonlinear-solid-mass} Nonlinear spring--dashpot system with an attached mass.]{\includegraphics[height=0.2\textwidth]{figures-explanation/nonlinear-solid-explicit-mass}}
  \caption{Nonlinear spring--dashpot system representing a generalisation of standard linear solid model.}
\end{figure}

In what follows we \emph{document the use of Colombeau algebra and the techniques developed by~\cite{prusa.v.rajagopal.kr:on} in the analysis of the response of specific nonlinear spring--dashpot systems}. In particular, we study a spring--dashpot system shown in Figure~\ref{fig:nonlinear-solid}, see Section~\ref{sec:gener-stand-line} for the results. Note that this spring dashpot--system can be interpreted as a generalisation of the standard linear solid model used in the description of the response of some viscoelastic materials, see for example~\cite{wineman.as.rajagopal.kr:mechanical}. Further, we study the same system, but with an attached mass as shown in Figure~\ref{fig:nonlinear-solid-mass}, see Section~\ref{sec:mass-spring-dashpot} for the results. In both cases we derive explicit formulae for the height of the jump in the response of the system to a step input.

\section{Preliminaries}
\label{sec:preliminaries}

We need a lemma derived by~\cite{prusa.v.rajagopal.kr:on}, see Lemma~\ref{lm:1} below\footnote{The numbering follows the work of~\cite{prusa.v.rajagopal.kr:on}.}. 
 The lemma deals with equation
 \begin{equation}
   \label{eq:75}
   h\Heaviside + \dd{}{t} \left(f\Heaviside + g\diracdelta\right) = 0,
 \end{equation}
where $\Heaviside$ denotes the Heaviside function, $\diracdelta$ is the Dirac distribution, $\diracdelta =_{\bydefinition} \dd{\Heaviside}{t}$, and $f$, $g$ and $h$ are smooth functions. 

The solution of the equation proceeds as follows. First, $\Heaviside$, $f$, $g$, $h$ and $\diracdelta$ are interpreted as the corresponding elements---the generalised functions---in Colombeau algebra. The corresponding elements are denoted as $\genfn{\Heaviside}$, $\genfn{f}$, $\genfn{g}$, $\genfn{h}$ and $\genfn{\diracdelta}$. Second, equality in~\eqref{eq:75} is interpreted as \emph{the equivalence in the sense of association} in Colombeau algebra. Doing so, one gets restrictions on the point values of $f$, $g$ and $h$, see the lemma. The reader interested in details is kindly referred to~\cite{prusa.v.rajagopal.kr:on}. 

\setcounter{theorem}{8}
\begin{lemma}
  \label{lm:1}
  Let $f$, $g$ and $h$ be continuously differentiable functions in $\R$. The generalised function $\genfn{h}\genfn{\Heaviside} + \dd{}{t} \left(\genfn{f}\genfn{\Heaviside} + \genfn{g}\genfn{\diracdelta}\right)$ vanishes in the sense of association, that is
  \begin{equation}
    \label{eq:7}
    \genfn{h}\genfn{\Heaviside} + \dd{}{t} \left(\genfn{f}\genfn{\Heaviside} + \genfn{g}\genfn{\diracdelta}\right) \approx \genfn{0},
  \end{equation}
  if and only if 
$      \left.
        f
      \right|_{t=0+}
      =
      0
$%
,
$
\left.
  g
\right|_{t=0+}
=
0
$%
,
and if the ordinary differential equation
$
  h 
  + 
  \dd{f}{t} 
  = 
  0
$%
, holds for all $t\geq 0$.
\end{lemma}


\section{Spring--dashpot system -- a generalisation of standard linear solid model}
\label{sec:gener-stand-line}
Let us now consider a specific spring--dashpot system namely the system shown in Figure~\ref{fig:nonlinear-solid}. The spring--dashpot system is composed of two springs and one dashpot. If all the components are linear and if the model is interpreted as model for the response of a viscoelastic material, then we would be working with the so called \emph{standard linear solid model}, see for example~\cite{wineman.as.rajagopal.kr:mechanical}. However, we assume that the components are \emph{nonlinear}, and that the stress--strain relations read
\begin{equation}
  \label{eq:12}
    \dd{\varepsilon_1}{t} = g_1(\sigma_1), \quad
    \varepsilon_2 = g_2(\sigma_2), \quad
    \sigma_3 = g_3(\varepsilon_3),
\end{equation}
where $\left\{\sigma_i\right\}_{i=1}^3$ and $\left\{\varepsilon \right\}_{i=1}^3$ denote the stress and strain in the respective element, and $g_1$, $g_2$ and $g_3$ are smooth invertible functions\footnote{Usually, the stress--strain relations are written in the form $\sigma_1 = h_1 \left( \dd{\varepsilon_1}{t} \right)$, $\sigma_2 = h_2(\varepsilon_2)$ and $\sigma_3 = g_3(\varepsilon_3)$, that is the stress is expressed as an explicit function of the strain or the strain rate. This is not the optimal way as how to handle constitutive relations, and the way of writing down the constitutive relations we have chosen in~\eqref{eq:12} might be in some situations the preferable one, see for example~\cite{rajagopal.kr:on*3,rajagopal.kr:generalized} and \cite{prazak.d.rajagopal.kr:mechanical}. But since we assume invertibility of functions $g_1$, $g_2$ and $g_3$ we can freely use either the classical way or the alternative way of writing the constitutive relations. The choice we make is convenient with respect to the ongoing calculations. 

Moreover, in the context of spring--dashpot systems the quantities $\left\{\sigma_i\right\}_{i=1}^3$  should be referred to as the forces and  quantities $\left\{\varepsilon \right\}_{i=1}^3$ should be referred to as the relative displacements, but we stick to the terminology that is used in the theory of viscoelasticity, and we shall refer to these quantities as stresses and strains.
} 
 such that $s=0$ implies $g_i(s)=0$ for all $i=1, \dots, 3$. Let us now find the relation between the total strain $\varepsilon$ and the stress $\sigma$ acting on the system.

The relation between the total strain $\varepsilon$ and the stress $\sigma$ can be derived by appealing to the standard procedure. If $A$ denotes the part that consists of the spring and the dashpot in series, see Figure~\ref{fig:nonlinear-solid}, then the stress--strain relation for this part of the system reads
  \begin{equation}
    \label{eq:13}
    g_1(\sigma_A) + \dd{}{t}g_2(\sigma_A) = \dd{\varepsilon_A}{t},
  \end{equation}
  where $\sigma_A$ is the stress in this part of the system. (The equation follows from equations $\sigma_A = \sigma_1 = \sigma_2$ and $\varepsilon_1 + \varepsilon_2 = \varepsilon_A$.) In the remaining part $B$ of the system we have the stress--strain relation
  $
  \sigma_B = g_3(\varepsilon_B)
  $.
  Consequently, appealing to relations $\sigma = \sigma_A + \sigma_B$ and $\varepsilon_A = \varepsilon_B = \varepsilon$, the sought global stress--strain relation reads
  \begin{equation}
    \label{eq:14}
    g_1(\sigma - g_3(\varepsilon)) + \dd{}{t} \Big[ g_2(\sigma - g_3(\varepsilon)) \Big]= \dd{\varepsilon}{t}.
  \end{equation}
Note that if the constitutive functions $g_i$ are chosen as $g_1(s)=\frac{1}{\mu_1} s$, $g_2(s)=\frac{1}{E_2}s$ and $g_3(s)=E_3 s$, where $\mu_1$, $E_1$ and $E_2$ are constants, then~\eqref{eq:14} reduces to 
$
    \frac{1}{E_2}\dd{\sigma}{t}
    +
    \frac{\sigma}{\mu_1}
    =
    \frac{E_3}{\mu_1}\varepsilon
    +
    \left(
      1
      +
      \frac{E_3}{E_2}
    \right)
    \dd{\varepsilon}{t}
$%
, which is the standard formula known form the linear setting.

Now the problem of interest is the response of the system to the input in the form $\varepsilon(t) = \varepsilon_+(t)\Heaviside(t)$, where $\varepsilon_+$ is a \emph{given} smooth function. The response $\sigma(t)$ takes the form $\sigma(t) = \sigma_+(t) \Heaviside(t)$, where $\sigma_+$ is a smooth function. In virtue of Lemma~\ref{lm:1} we see that the algebraic equation relating the unknown jump in the response $\left. \sigma_+ \right|_{t=0+}$  to the known jump $\left. \varepsilon_+ \right|_{t=0+}$  in the input reads
  \begin{equation}
    \label{eq:15}
    g_2\left(\left. \sigma_+ \right|_{t=0+} - g_3 \left( \left. \varepsilon_+ \right|_{t=0+} \right) \right)
    -
    \left. \varepsilon_+ \right|_{t=0+}
    =
    0
    .
  \end{equation}
  Further, the stress--strain relation~\eqref{eq:14} holds everywhere except at the jump. In particular $\sigma_+$ solves for $t>0$ the differential equation
  \begin{equation}
    \label{eq:69}
    g_1(\sigma_+ - g_3(\varepsilon_+)) + \dd{}{t} \Big[ g_2(\sigma_+ - g_3(\varepsilon_+)) \Big]= \dd{\varepsilon_+}{t},
  \end{equation}
  where solution to~\eqref{eq:15} determines the initial condition $\left. \sigma_+ \right|_{t=0+}$. Naive justification of the result is based on rewriting~\eqref{eq:14} as 
\begin{equation}
  \label{eq:68}
  g_1(\sigma - g_3(\varepsilon)) + \dd{}{t} \Big[ g_2(\sigma - g_3(\varepsilon)) - \varepsilon \Big]= 0,
\end{equation}
which upon substituting formulae $\varepsilon(t) = \varepsilon_+(t)\Heaviside(t)$ and $\sigma(t) = \sigma_+(t) \Heaviside(t)$ yields
  \begin{equation}
    \label{eq:16}
    g_1(\sigma_+ - g_3(\varepsilon_+))\Heaviside + \dd{}{t} \Big[ \left( g_2(\sigma_+ - g_3(\varepsilon_+)) - \varepsilon_+ \right) \Heaviside \Big]= 0,
  \end{equation}
  where we have used the fact that $g_i(s \Heaviside(s))=g_i(s)\Heaviside(s)$ for $i=1\dots, 3$. (This observation follows from the chosen form of the constitutive relations.) Equation~\eqref{eq:16} has the form discussed in Lemma~\ref{lm:1}, and the jump condition~\eqref{eq:15} follows immediately. 

Note that if we were using for example $g_1(s) =_{\bydefinition} (1 + s)^n s$, where $n \in \N$, then the equality $g_1(s \Heaviside(s)) = g_1(s) \Heaviside(s)$ we have used above, would bring us dangerously close to the equality $\Heaviside^n = \Heaviside$. However, using such equality could lead to paradoxical results, see the introduction, and extreme caution should be exercised. \emph{A rigorous justification of the manipulation used above is extremely desirable.} The justification of the procedure the setting of Colombeau algebra follows from the work of \cite{prusa.v.rajagopal.kr:on}.

\section{Mass--spring--dashpot system}
\label{sec:mass-spring-dashpot}
Let us now investigate the response of a more complex system. If a mass $m$ is attached to the spring--dashpot system, see Figure~\ref{fig:nonlinear-solid-mass}, then the time evolution of position $x$ of the mass $m$ is described by nonlinear ordinary differential equations
\begin{subequations}
  \label{eq:17}
  \begin{align}
    \label{eq:18}
    m \ddd{x}{t} &= F - \sigma, \\
    \label{eq:19}
    g_1(\sigma - g_3(\varepsilon)) + \dd{}{t} g_2(\sigma - g_3(\varepsilon)) &= \dd{\varepsilon}{t},
  \end{align}
\end{subequations}
where $F$ denotes the external force, $\varepsilon$ is the strain $\varepsilon =_{\bydefinition} \frac{x - x_{\equilibrium}}{x_{\equilibrium}}$, and $x_{\equilibrium}$ denotes the equilibrium length.

\subsection{Response to a step input}
\label{sec:response-input-with}
Let us now assume that the time evolution of the position $x$ is \emph{given} as
\begin{equation}
  \label{eq:20}
  x = x_{\equilibrium} + (x_+ - x_{\equilibrium}) \Heaviside,
\end{equation}
where $x_+$ is a known smooth function of time. This means that the mass $m$ is suddenly moved from the equilibrium position $x_{\equilibrium}$ to the position $\left. x_+ \right|_{t=0+}$, hence we are again dealing with a step input\footnote{In practice the mass can not suddenly jump from one place to the other. But if the mass moves sufficiently fast---compared to the observation time of the system---then it makes sense to model its motion as a sudden jump, see \cite{prusa.rajagopal.kr:jump} for the discussion.}
. The task is to find the force $F$ that corresponds to such a motion/input. 

The sought force $F$ is interpreted as a generalised function $\genfn{F}$, and it is assumed to take the form
\begin{equation}
  \label{eq:21}
  \genfn{F} =_{\bydefinition} \dd{}{t} \left( f_+ \genfn{\Heaviside} + g_+ \genfn{\diracdelta} \right),
\end{equation}
where $f_+$ and $g_+$ are smooth functions\footnote{The \emph{ansatz} can be found by experimenting with a general \emph{ansatz} $\genfn{F} =_{\bydefinition} u \genfn{\Heaviside} + v \genfn{\diracdelta} + w \dd{\genfn{\diracdelta}}{t}$. Manipulating the general~\emph{ansatz} one quickly finds that the general \emph{ansatz} must be of the special form~\eqref{eq:21}, otherwise there is no chance to use Lemma~\ref{lm:1} that guarantees the solution being a generalised function that is associated to some classical distribution. Note that the counterpart of the Dirac distribution in Colombeau algebra is defined in the same manner as in the classical case, that is $\genfn{\diracdelta}=_{\bydefinition}\dd{\genfn{\Heaviside}}{t}$.} that need to be found. Similarly, function~\eqref{eq:20} describing the time evolution of the position $x$ is understood as a generalised function
\begin{equation}
  \label{eq:22}
  \genfn{x} = x_{\equilibrium} + (x_+ - x_{\equilibrium}) \genfn{\Heaviside}.
\end{equation}
The core of the problem is to find the values of $f_+$ and $g_+$ at $t=0+$. As we shall see completing this task requires one to find the stress~$\sigma$ in the spring--dashpot system at time $t=0+$, which is the problem that has been studied in the previous section.

In order to solve the new problem, we can proceed as follows. Substituting~\eqref{eq:22} and~\eqref{eq:21} into governing equations~\eqref{eq:17} with the equalities interpreted as the equivalences in the sense of association yields
\begin{subequations}
  \label{eq:23}
  \begin{align}
    \label{eq:24}
    m \dd{}{t} \left[ \dd{x_+}{t}\genfn{\Heaviside} + (x_+ - x_{\equilibrium}) \genfn{\diracdelta} \right] &\approx \dd{}{t}\left[ f_+ \genfn{\Heaviside} + g_+ \genfn{\diracdelta} \right] - \genfn{\sigma}, \\
    \label{eq:25}
    g_1(\genfn{\sigma} - g_3(\genfn{\varepsilon})) + \dd{}{t} g_2(\genfn{\sigma} - g_3(\genfn{\varepsilon})) &\approx \dd{\genfn{\varepsilon}}{t},
  \end{align}
\end{subequations}
where the strain $\genfn{\varepsilon}$ is interpreted as the generalised function
$
  \genfn{\varepsilon} =  \varepsilon_+ \genfn{\Heaviside}
$,
where
$
\varepsilon_+ =_{\bydefinition}
\frac{x_+ - x_{\equilibrium}}{x_{\equilibrium}}
$.

Concerning the solution to~\eqref{eq:25} we can use the results obtained in the previous section. We know that $\genfn{\sigma}$ is given by the formula
$
  \genfn{\sigma} = \sigma_+ \genfn{\Heaviside}
$,
where $\sigma_+$ is for $t>0$ the solution to the nonlinear ordinary differential equation
\begin{subequations}
  \label{eq:26}
  \begin{align}
    \label{eq:27}
     g_1(\sigma_+ - g_3(\varepsilon_+)) + \dd{}{t} g_2(\sigma_+ - g_3(\varepsilon_+)) &= \dd{\varepsilon_+}{t}, \\
     \label{eq:28}
     \left. \sigma_+ \right|_{t=0+} &= \sigma_0,
  \end{align}
\end{subequations}
and the initial condition~\eqref{eq:28} is obtained as the solution to the algebraic equation
\begin{equation}
  \label{eq:29}
  g_2(\sigma_0 - g_3(\left. \varepsilon_+ \right|_{t=0+})) = \left. \varepsilon_+ \right|_{t=0+},
\end{equation}
see condition~\eqref{eq:15}. This means that $\genfn{\sigma}$ is determined by~$\genfn{\varepsilon}$, and it can be treated as a known function in~\eqref{eq:24}. 

Rearranging the terms in~\eqref{eq:24} and utilising the fact that $\genfn{\sigma} = \sigma_+ \genfn{\Heaviside}$ is a known function yields the equation
\begin{equation}
  \label{eq:30}
   \sigma_+\genfn{\Heaviside} + \dd{}{t} \left[ \left( m\dd{x_+}{t} - f_+ \right)\genfn{\Heaviside} + \left( m(x_+ - x_{\equilibrium}) -  g_+ \right) \genfn{\diracdelta} \right] \approx \genfn{0}.
\end{equation}
The equation takes the form analysed in Lemma~\ref{lm:1}, and the lemma gives us two pointwise conditions that has to hold at $t=0+$,
\begin{subequations}
  \label{eq:31}
  \begin{align}
    \label{eq:32}
    \left. \left( m(x_+ - x_{\equilibrium}) -  g_+ \right) \right|_{t=0+} &= 0, \\ 
    \label{eq:33}
    \left. \left( m\dd{x_+}{t} - f_+ \right) \right|_{t=0+} &= 0.
  \end{align}
\end{subequations}
These conditions fix the value of the unknown function $g_+$ at $t=0+$ and the value of $f_+$ at $t=0+$ in terms of the problem data, that is in terms of the given function $x_+$. The last condition in Lemma~\ref{lm:1} yields the differential equation for the function $f_+$,
\begin{equation}
  \label{eq:34}
   m\ddd{x_+}{t} =  \dd{f_+}{t} - \sigma_+,
\end{equation}
which must be solved subject to the initial condition~\eqref{eq:33}. Solution to~\eqref{eq:34} then determines the function $f_+$ in the \emph{ansatz}~\eqref{eq:21}.

\subsection{Summary}
\label{sec:summary}
We can therefore conclude that the force response of the system shown in Figure~\ref{fig:nonlinear-solid-mass} to the prescribed step input 
\begin{equation}
  \label{eq:35}
  \genfn{x} = x_{\equilibrium} + (x_+ - x_{\equilibrium}) \genfn{\Heaviside},
\end{equation}
where~$x_+$ is a smooth function, is given by the formula 
\begin{equation}
  \label{eq:36}
  \genfn{F} = \dd{}{t} \left( f_+ \genfn{\Heaviside} + g_+ \genfn{\diracdelta} \right),
\end{equation}
where function $f_+$ is for $t>0$ the solution to the system of nonlinear ordinary differential equations
\begin{subequations}
  \label{eq:complete-solution-spring-mass-dashpot}
  \begin{align}
    \label{eq:37}
    m\ddd{x_+}{t} &=  \dd{f_+}{t} - \sigma_+, \\
    \label{eq:38}
    g_1(\sigma_+ - g_3(\varepsilon_+)) + \dd{}{t} g_2(\sigma_+ - g_3(\varepsilon_+)) &= \dd{\varepsilon_+}{t},
  \end{align}
  for unknown functions $f_+$ and $\sigma_+$ with initial conditions
  \begin{align}
    \label{eq:39}
    \left. f_+ \right|_{t=0+} &= m \left. \dd{x_+}{t} \right|_{t=0+}, \\
    \label{eq:40}
    \left. \sigma_+ \right|_{t=0+} &= \inverse{g_2}\left( \left. \varepsilon_+ \right|_{t=0+} \right) + g_3 \left( \left. \varepsilon_+ \right|_{t=0+} \right),
  \end{align}
  where $\varepsilon_+ =_{\bydefinition} \frac{x_+ - x_{\equilibrium}}{x_{\equilibrium}}$ denotes the total strain, $\inverse{g_2}$ stands for the inverse of function $g_2$ in the constitutive relation~\eqref{eq:12}, and function $g_+$ is a smooth function that satisfies the condition
  \begin{equation}
    \label{eq:41}
     \left. g_+ \right|_{t=0+} =   m( \left. x_+ \right|_{t=0+} - x_{\equilibrium}).  
  \end{equation}
\end{subequations}

Note that the \emph{solution works even if we invert the role of the input and the response}. Indeed, if $f_+$ and $g_+$ are given, then~\eqref{eq:complete-solution-spring-mass-dashpot} is a system of equations that allows one to determine function $x_+$ in the formula~\eqref{eq:35} for the motion caused by the force~\eqref{eq:36}.

\section{Sequential interpretation of the result and numerical experiments}
\label{sec:numer-exper}
Let us now document the utility of the derived formulae in a practical application. Let us assume that the task is to determine the force $F$ such that the position of the mass $m$ in the system shown in Figure~\ref{fig:nonlinear-solid} is given by the formula
\begin{equation}
  \label{eq:42}
  x = x_{\equilibrium} + (x_{\final} - x_{\equilibrium}) \Heaviside,
\end{equation}
where $x_{\final}$ is a fixed \emph{constant}. This means that we want the mass to instantaneously move from the equilibrium position $x_{\equilibrium}$ to a new position $x_{\final}$, and then we want the mass to stay at rest at the new position. We have shown, see the previous section, that it is possible to explicitly write down the solution to this fictitious control problem, provided that we work in the setting of Colombeau algebra.

Moreover, Heaviside function $\genfn{\Heaviside}$ as an element in Colombeau algebra can be seen as a ``cluster'' of smooth functions that approximate the piecewise constant function $\Heaviside$ as defined in~\eqref{eq:1}, see for example~\cite{prusa.v.rajagopal.kr:on} for details. Let us now exploit this ``sequential'' interpretation of the elements in Colombeau algebra. We define the sequence of functions
\begin{equation}
  \label{eq:71}
  H_n
  =_{\bydefinition}
  \begin{cases}
    0, & t \in \left( - \infty, 0 \right), \\
    t^4(d_3 t^3 + d_2 t^2 + d_1t + d_0) , &t \in \left[ 0, \frac{1}{n} \right], \\
    1, &t \in \left( \frac{1}{n}, +\infty \right),
  \end{cases}
\end{equation}
where $d_3=-20n^7$, $d_2=70n^6$, $d_1=-84n^5$ and $d_0=35n^4$, that for large $n\in \N$ recovers the exact Heaviside function $\Heaviside$. Further, the sequence of forces
\begin{equation}
  \label{eq:72}
  F_n =_{\bydefinition}
  \dd{}{t} \left( f_+ \Heaviside_n + g_+ \dd{\Heaviside_n}{t} \right)
\end{equation}
approximates the exact force $\genfn{F}$ given by the formula~\eqref{eq:36}, provided that $f_+$ and $g_+$ are calculated from the given $x_{\final}$ by the formulae derived in Section~\ref{sec:summary}. Since function $H_n$ is for any $n\in \N$ a continuous function and it has continuous first and second derivative, we see that the differential equation~\eqref{eq:17} has the \emph{classical} solution for any approximate force $F_n$. 

Let us denote $x_n$ the solution of~\eqref{eq:17} corresponding to the approximate force $F_n$. The solution can be found either explicitly---which is not feasible except of some special cases---or numerically using modern solvers for stiff differential equations. The sequence~$x_n$ obtained as the sequence of responses to the approximated force inputs $F_n$ should for large $n$ recover the desired exact response~\eqref{eq:42}. As we shall demonstrate below by a \emph{numerical experiment}, this is indeed the case.

\subsection{Specific constitutive relations for the numerical experiment}
\label{sec:spec-const-relat}
In order to do the numerical computations we need to fix the constitutive relations. We set
\begin{equation}
  \label{eq:48}
    g_1(s) =_{\bydefinition} \frac{1}{\alpha \mu_1} \left(\exponential{\alpha s} - 1\right), \quad
    g_2(s) =_{\bydefinition} \frac{1}{\beta E_2} \left(\exponential{\beta s} - 1\right),\quad
    g_3(s) =_{\bydefinition} E_3 \left(1 + \gamma s^2 \right)s,
\end{equation}
where $\alpha$, $\mu_1$, $E_2$, $\beta$, $E_3$ and $\gamma$ are positive constants. 

\subsection{Force necessary to cause the piecewise constant deformation -- analytical solution}
\label{sec:force-resp-piec}
We see that the inverse to $g_2$ reads $g_2^{-1}(s) = \frac{1}{\beta} \ln \left( 1 + \beta E_2 s\right)$, and from~\eqref{eq:40} we get an explicit formula for $\sigma_{\final} =_{\bydefinition} \left. \sigma_+ \right|_{t=0+}$, 
\begin{equation}
  \label{eq:49}
  \left. \sigma_+ \right|_{t=0+} =  \frac{1}{\beta} \ln \left( 1 + \beta E_2 \varepsilon_{\final} \right) + E_3 \left(1 + \gamma \varepsilon_{\final}^2 \right) \varepsilon_{\final}.
\end{equation}
Further, if we substitute the specific constitutive relations~\eqref{eq:48} into~\eqref{eq:38}, we see that~\eqref{eq:38} can be rewritten as
\begin{equation}
  \label{eq:50}
  \frac{1}{\alpha \mu_1} \left(\exponential{u_+} -1 \right) + \frac{1}{\beta E_2} \dd{}{t} \left(\exponential{\frac{\beta}{\alpha}u_+} -1 \right) = 0,
\end{equation}
where $u_+ =_{\bydefinition}  \alpha(\sigma_+ - g_3(\varepsilon_{\final}))$, and the initial condition for~\eqref{eq:50} reads $\left. u_+ \right|_{t=0+} = \frac{\alpha}{\beta} \ln \left( 1 + \beta E_2 \varepsilon_{\final} \right)$. (The initial condition follows from the definition of $u_+$ and the initial condition~\eqref{eq:49}.) Equation~\eqref{eq:50} can be solved explicitly provided that $\alpha = \beta$, \emph{which is the case we shall study in the rest of the section}. If $\alpha = \beta$, then the solution to~\eqref{eq:50} reads
$
  u_+ = \ln (1 + \left(\exponential{\left. u_+ \right|_{t=0+}} - 1\right) \exponential{-\frac{E_2}{\mu_1}t} )
$,
and going back to the original unknown $\sigma_+$ yields the sought explicit formula for~$\sigma_+$, 
\begin{equation}
  \label{eq:51}
  \sigma_+ = \frac{1}{\alpha}\ln \left(1 + \left(\exponential{\ln \left( 1 + \beta E_2 \varepsilon_{\final} \right)} - 1\right) \exponential{-\frac{E_2}{\mu_1}t}\right) + g_3(\varepsilon_{\final}).
\end{equation}
Note that $\sigma(t) \to g_3(\varepsilon_{\final})$ as $t \to +\infty$. This is an expected result from the physical point of view. If nothing moves, then the stress is controlled exclusively by the part B of the system, see Figure~\ref{fig:nonlinear-solid-mass}.

Having solved the equation for $\sigma_+$, we are ready to solve~\eqref{eq:37} for $f_+$. Since $x_+$ is in our case a constant $x_+ =_{\bydefinition} x_{\final}$, we see that the solution to~\eqref{eq:37} subject to initial condition~\eqref{eq:39} is 
\begin{equation}
  \label{eq:67}
  f_+ = \int_{s=0}^t \sigma_+(s) \, \diff s.
\end{equation}
Finally, function $g_+$ is given by the equation~\eqref{eq:41}, hence 
\begin{equation}
  \label{eq:73}
  g_+ = m (x_{\final} - x_{\equilibrium}).
\end{equation}

Having found $f_+$ and $g_+$, we can substitute into the \emph{ansatz} \eqref{eq:36}. Before doing so, we rewrite~\eqref{eq:36} as
$
  \genfn{F} 
  =  \dd{f_+}{t} \genfn{\Heaviside} + f_+ \genfn{\diracdelta} + \dd{}{t} \left( g_+ \dd{\genfn{\Heaviside}}{t} \right)
$%
,
that in virtue of~\eqref{eq:67} reduces to
\begin{equation}
  \label{eq:70}
  \genfn{F}
  \approx
  \sigma_+ \genfn{\Heaviside} + \dd{}{t} \left( g_+ \dd{\genfn{\Heaviside}}{t} \right).
\end{equation}
Here we have used the fact that $f_+ \genfn{\diracdelta} \approx \genfn{0}$ that follows from\footnote{Note that $f_+ \genfn{\diracdelta} \not = \genfn{0}$ if the equality is understood as the \emph{strict} equality in Colombeau algebra. This is one of the substantial differences between the classical theory of distributions and Colombeau algebra. In the classical theory one \emph{has} $f \diracdelta = 0$, provided that $f$ is a smooth function vanishing at zero. However, since the governing differential equations are formulated in terms of \emph{the equality in the sense of association}, then we can use $f_+ \genfn{\diracdelta} \approx \genfn{0}$, which \emph{is} true in Colombeau algebra, see for example~\cite{prusa.v.rajagopal.kr:on} for detailed discussion.} 
 equality $f_+(0)=0$.
Finally, substituting the explicit formulae for $\sigma_+$ and~$g_+$, see~\eqref{eq:51} and~\eqref{eq:73}, into~\eqref{eq:70} yields
\begin{equation}
  \label{eq:53}
  \genfn{F} 
  \approx
  \left[ 
    \frac{1}{\alpha}\ln \left(1 + \beta E_2 \varepsilon_{\final} \exponential{-\frac{E_2}{\mu_1}t} \right) 
    + 
    E_3 \left(1 + \gamma \varepsilon_{\final}^2 \right) \varepsilon_{\final}
  \right] \genfn{\Heaviside}
  +
  \dd{}{t}
  \left(
    m (x_{\final} - x_{\equilibrium})
    \dd{\genfn{\Heaviside}}{t}
  \right)
  .
\end{equation}

This result corresponds to the intuition we have on the behaviour of such a simple system. The force necessary to instantaneously move the mass from one place to the other is composed of two parts. 

The first contribution to the total force originates in the necessity to instantaneously deform the spring--dashpot system. However, the instantaneous deformation of the system is the \emph{elastic} one, meaning that the elastic elements---the springs---are the only active elements in the instantaneous response. Indeed, at $t=0+$ the response is given by~\eqref{eq:49}, which is in fact the sum of the stresses in the springs $\left. \sigma_+ \right|_{t=0+} = \left. \sigma_2 \right|_{t=0+} + \left. \sigma_3  \right|_{t=0+} = \inverse{g_2}(\varepsilon_{\final}) + g_3(\varepsilon_{\final})$. Note that if $\alpha=\beta \to 0+$ and $\gamma \to 0+$, that is if the springs are linear springs, then $\left. \sigma_+ \right|_{t=0+} = \left(E_2 + E_1\right) \varepsilon_{\final}$, which is the standard result obtained in the linear setting.

The second contribution to the total force comes from the fact that one needs infinite acceleration to initiate the motion, and the acceleration must be straight away followed by an infinite deceleration to instantaneously stop the motion at the right place. This contribution is captured by the term  
$
\dd{}{t}
\left(
  m (x_{\final} - x_{\equilibrium})
  \dd{\genfn{\Heaviside}}{t}
\right)
\approx
m (x_{\final} - x_{\equilibrium}) \dd{\genfn{\diracdelta}}{t}
$, %
and it is exclusively due to the \emph{inertia} of the system.

\subsection{Numerical solution}
\label{sec:numerical-solution}

\begin{figure}[h]
  \centering
  \subfloat[\label{fig:input-response-mass--spring--dashpot-force}Approximated input, external applied force $F_n$.]{\includegraphics[width=0.46\textwidth]{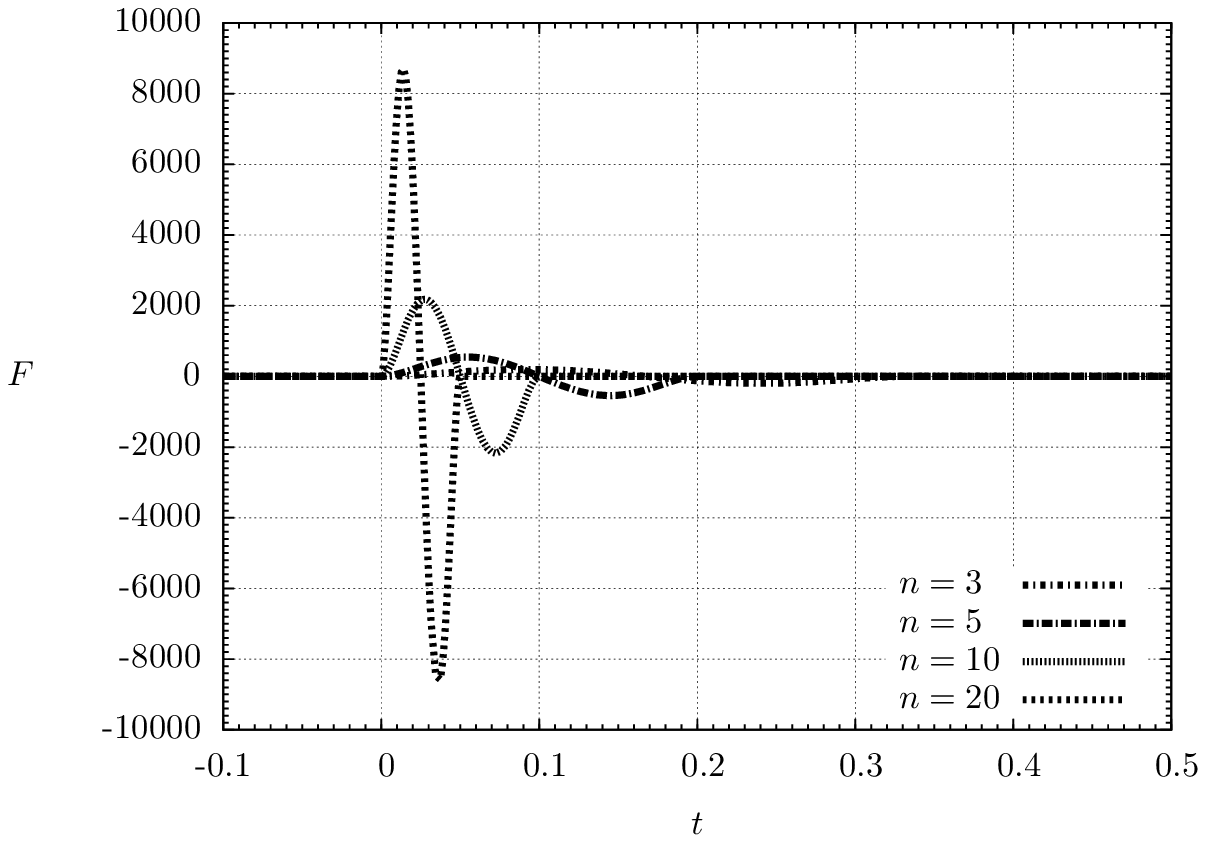}}
  \qquad
  \subfloat[\label{fig:input-response-mass--spring--dashpot-position}Response to input $F_n$, numerical solution, position $x_n$.]{\includegraphics[width=0.46\textwidth]{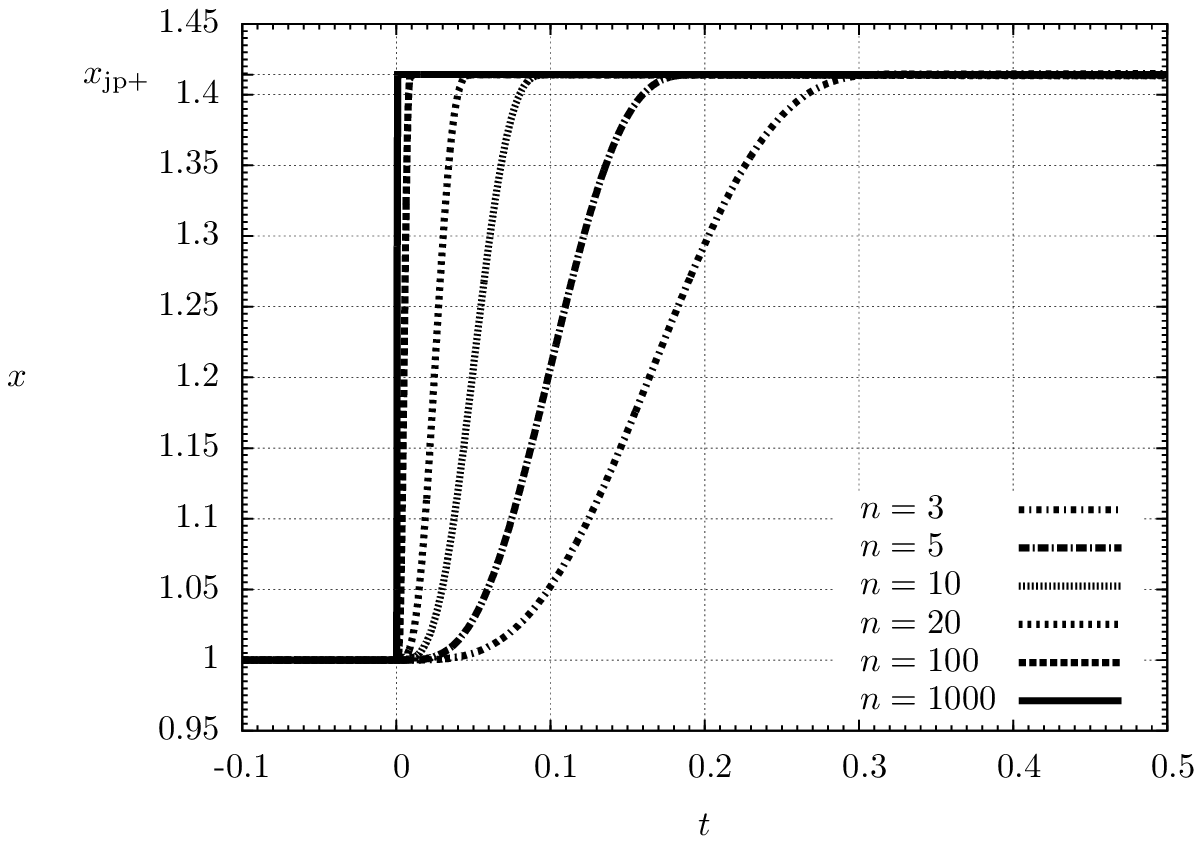}}
  \\
  \subfloat[\label{fig:input-response-mass--spring--dashpot-velocity}Response to input $F_n$, numerical solution, velocity $v_n$.]{\includegraphics[width=0.46\textwidth]{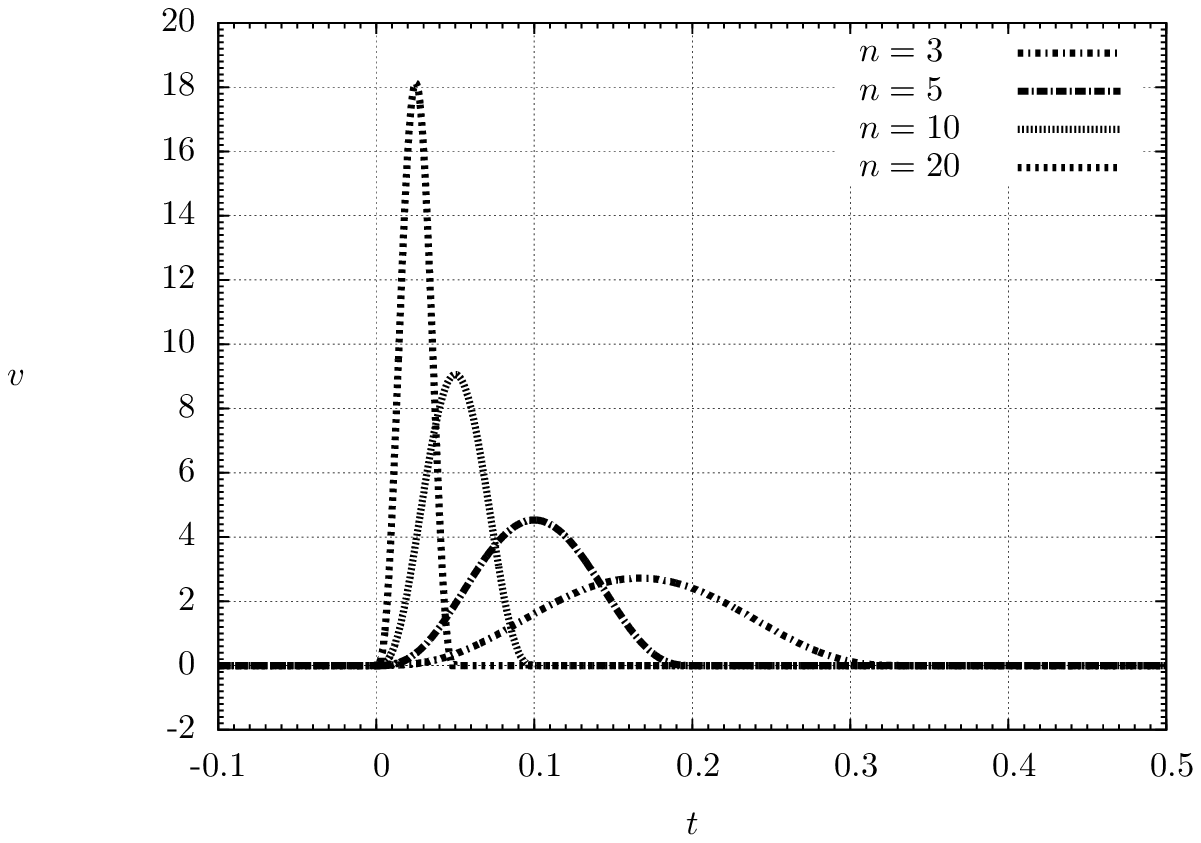}}
  \qquad
  \subfloat[\label{fig:input-response-mass--spring--dashpot-stress}Response to input $F_n$, numerical solution, stress $\sigma_n$.]{\includegraphics[width=0.46\textwidth]{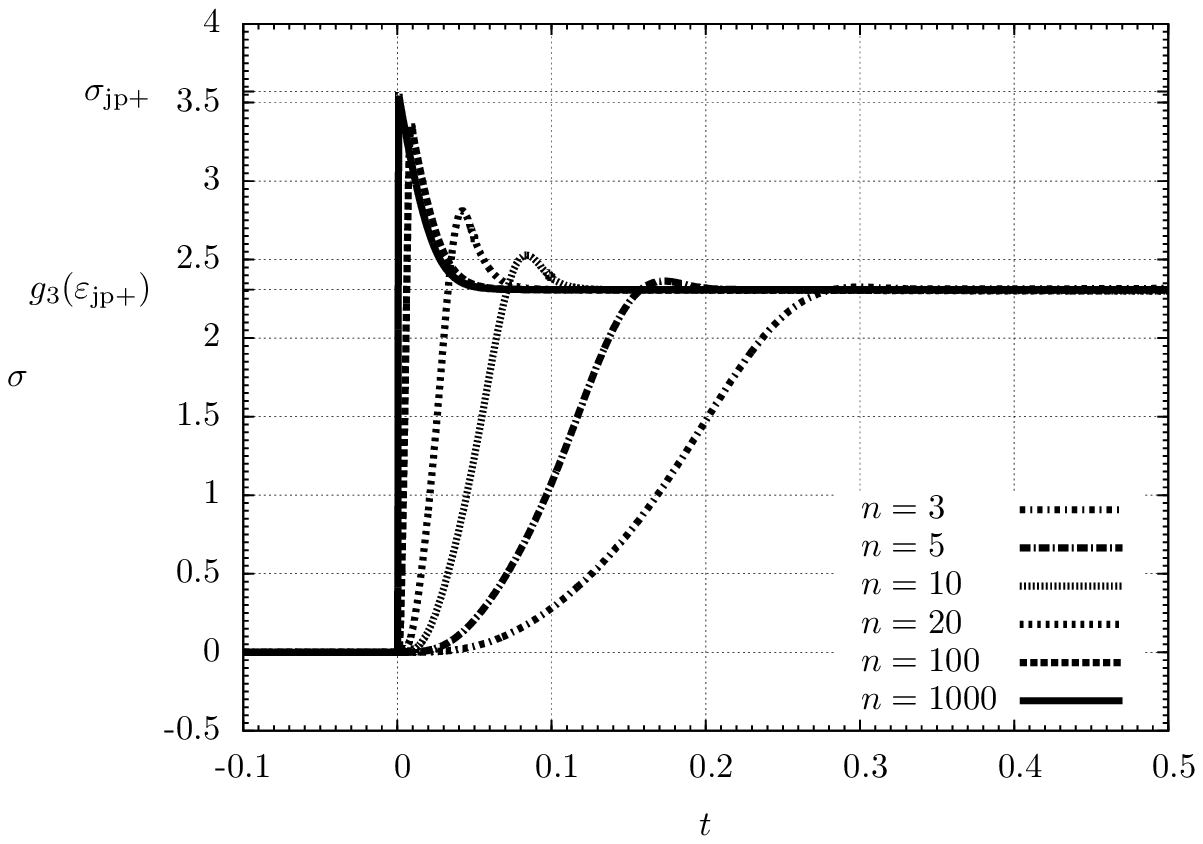}}
  \caption{Response of a mass--spring--dashpot system. Numerical solution for a sequence of approximated inputs $F_n$, see~\eqref{eq:76}. System parameters are $x_{\equilibrium}=1$, $m=7$,  $\alpha=\beta=\sqrt{3}$, $\gamma=5$, $E_1=11$, $E_2=3$, $\mu_1=\frac{1}{10}$. Expected height of the jump in $\sigma$ is $\sigma_{\final} = 3.570244804 \dots$, expected response in position $x$ is the step change from $x_{\equilibrium}=1$ to $x_{\final}= \sqrt{2}$. Solved in MAPLE using adaptive Runge--Kutta--Fehlberg method \texttt{rkf45}.}
  \label{fig:input-response-mass--spring--dashpot}
\end{figure}

Let us now take the sequence 
\begin{equation}
  \label{eq:76}
  F_n =   
  \left[ 
    \frac{1}{\alpha}\ln \left(1 + \beta E_2 \varepsilon_{\final} \exponential{-\frac{E_2}{\mu}t} \right) 
    + 
    E_3 \left(1 + \gamma \varepsilon_{\final}^2 \right) \varepsilon_{\final}
  \right] \Heaviside_n
  +
  \dd{}{t}
  \left(
    m (x_{\final} - x_{\equilibrium})
    \dd{\Heaviside_n}{t}
  \right)
\end{equation}
of \emph{continuous} functions $F_n$ approximating the exact force~\eqref{eq:53} that leads to the step change~\eqref{eq:42} in the position $x$. Some members of the sequence of the approximated forces is shown in Figure~\ref{fig:input-response-mass--spring--dashpot-force}. Further, let us find numerically the sequence of the functions $x_n$ that correspond to the sequence of approximated forces $F_n$. The numerical solution to~\eqref{eq:17} is obtained by solving the equivalent first order system
\begin{subequations}
  \label{eq:62}
  \begin{align}
    \label{eq:63}
    \dd{v}{t} &= \frac{F - \sigma}{m}, \\
    \label{eq:64}
    \dd{x}{t} &= v, \\
    \label{eq:65}
    \dd{\sigma}{t}
    &=
    \frac{
      \frac{v}{x_{\equilibrium}}
      -
      g_1\left(\sigma -  g_3\left( \frac{x - x_{\equilibrium}}{x_{\equilibrium}} \right) \right)
    }
    {
      \left. \dd{g_2}{s} \right|_{s = \sigma  - g_3\left( \frac{x - x_{\equilibrium}}{x_{\equilibrium}} \right)}
    }
    +
    \left.
      \dd{g_3}{s}
    \right|_{s = \frac{x - x_{\equilibrium}}{x_{\equilibrium}}}
    \frac{v}{x_{\equilibrium}}
    ,
  \end{align}
\end{subequations}
for the triple $x_n$, $v_n$ and $\sigma_n$, and it is plotted for various values of $n$ in Figure~\ref{fig:input-response-mass--spring--dashpot}. Response $x_n$, $v_n$ and $\sigma_n$ to the approximated input $F_n$ indeed approaches for large values of $n$ the exact response predicted by the theory. In particular, the response $x_n$ tends to the desired step response~\eqref{eq:42}, see Figure~\ref{fig:input-response-mass--spring--dashpot-position}, and the stress sequence $\sigma_n$ recovers the jump response with the predicted jump height~$\sigma_{\final}$, see Figure~\ref{fig:input-response-mass--spring--dashpot-stress}.

The lesson learned from the numerical experiment is that \emph{extremely fast but smooth changes} can be modelled as changes with jump discontinuities and \emph{vice versa}. Indeed, the fast changes---large values of $n$---are virtually indistinguishable from the step change. The benefit of using Colombeau algebra is that the theory guarantees such correspondence for some \emph{nonlinear} systems. (Note, however, that there exist nonlinear systems where the jump in the response is sensitive to the particular way of smoothing the jump in the input, see~\cite{prusa.rajagopal.kr:jump} for details.) Moreover, the theory provides an \emph{exact characterisation} of the behaviour at the jump discontinuity. Such result can not be obtained on the basis of numerical calculations.

\section{Conclusion}
\label{sec:conclusion}
Colombeau algebra is an extension of the classical theory of distributions into the nonlinear setting. Despite its apparent complexity, Colombeau algebra can be used, as shown above, in symbolic calculations with almost the same ease as the classical theory of distributions. In particular, Colombeau algebra provides one a concept of solution to a nonlinear ordinary differential equation with jump discontinuities, and it allows one to explicitly characterise the behaviour of the solution at the point of jump discontinuity. The existence of a relatively easy to handle nonlinear theory of distributions opens up the possibility to analyse the response of various systems governed by nonlinear ordinary differential equations to inputs with jump discontinuities.


\bibliographystyle{chicago}
\bibliography{vit-prusa}

\begin{thebibliography}{}

\bibitem[\protect\citeauthoryear{Colombeau}{Colombeau}{1984}]{colombeau.j:new}
Colombeau, J.-F. (1984).
\newblock {\em New generalized functions and multiplication of distributions},
  Volume~84 of {\em North-Holland Mathematics Studies}.
\newblock Amsterdam: North-Holland Publishing Co.
\newblock Notas de Matem{\'a}tica [Mathematical Notes], 90.

\bibitem[\protect\citeauthoryear{Colombeau}{Colombeau}{1992}]{colombeau.j:multiplication}
Colombeau, J.-F. (1992).
\newblock {\em Multiplication of distributions}, Volume 1532 of {\em Lecture
  Notes in Mathematics}.
\newblock Berlin: Springer-Verlag.
\newblock A tool in mathematics, numerical engineering and theoretical physics.

\bibitem[\protect\citeauthoryear{Pra\v{z}\'ak and Rajagopal}{Pra\v{z}\'ak and
  Rajagopal}{2012}]{prazak.d.rajagopal.kr:mechanical}
Pra\v{z}\'ak, D. and K.~R. Rajagopal (2012).
\newblock Mechanical oscillators described by a system of
  differential-algebraic equations.
\newblock {\em Applications of Mathematics\/}~{\em 57\/}(2), 129--142.

\bibitem[\protect\citeauthoryear{Pr\r{u}\v{s}a and Rajagopal}{Pr\r{u}\v{s}a and
  Rajagopal}{2011}]{prusa.rajagopal.kr:jump}
Pr\r{u}\v{s}a, V. and K.~R. Rajagopal (2011).
\newblock Jump conditions in stress relaxation and creep experiments of
  {B}urgers type fluids: {A} study in the application of {C}olombeau algebra of
  generalized functions.
\newblock {\em Z. Angew. Math. Phys.\/}~{\em 62\/}(4), 707--740.

\bibitem[\protect\citeauthoryear{Pr\r{u}\v{s}a and Rajagopal}{Pr\r{u}\v{s}a and
  Rajagopal}{2016}]{prusa.v.rajagopal.kr:on}
Pr\r{u}\v{s}a, V. and K.~R. Rajagopal (2016).
\newblock On the response of physical systems governed by nonlinear ordinary
  differential equations to step input.
\newblock {\em Int. J. Non-Linear Mech.\/}~{\em 81}, 207--221.

\bibitem[\protect\citeauthoryear{Rajagopal}{Rajagopal}{2003}]{rajagopal.kr:on*3}
Rajagopal, K.~R. (2003).
\newblock On implicit constitutive theories.
\newblock {\em Appl. Math.\/}~{\em 48\/}(4), 279--319.

\bibitem[\protect\citeauthoryear{Rajagopal}{Rajagopal}{2010}]{rajagopal.kr:generalized}
Rajagopal, K.~R. (2010).
\newblock A generalized framework for studying the vibrations of lumped
  parameter systems.
\newblock {\em Mech. Res. Commun.\/}~{\em 37\/}(5), 463--466.

\bibitem[\protect\citeauthoryear{Rosinger}{Rosinger}{1987}]{rosinger.ee:generalized}
Rosinger, E.~E. (1987).
\newblock {\em Generalized solutions of nonlinear partial differential
  equations}, Volume 146 of {\em North-Holland Mathematics Studies}.
\newblock Amsterdam: North-Holland Publishing Co.
\newblock Notas de Matem{\'a}tica [Mathematical Notes], 119.

\bibitem[\protect\citeauthoryear{Rosinger}{Rosinger}{1990}]{rosinger.ee:nonlinear}
Rosinger, E.~E. (1990).
\newblock {\em Nonlinear partial differential equations: An algebraic view of
  generalized solutions}, Volume 164 of {\em North-Holland Mathematics
  Studies}.
\newblock Amsterdam: North-Holland Publishing Co.

\bibitem[\protect\citeauthoryear{Schwartz}{Schwartz}{1954}]{schwartz.l:sur}
Schwartz, L. (1954).
\newblock Sur l'impossibilit\'e de la multiplication des distributions.
\newblock {\em C. R. Acad. Sci. Paris\/}~{\em 239}, 847--848.

\bibitem[\protect\citeauthoryear{Schwartz}{Schwartz}{1966}]{schwartz.l:theorie}
Schwartz, L. (1966).
\newblock {\em Th\'eorie des distributions}.
\newblock Publications de l'Institut de Math\'ematique de l'Universit\'e de
  Strasbourg, No. IX-X. Nouvelle \'edition, enti\'erement corrig\'ee, refondue
  et augment\'ee. Hermann, Paris.

\bibitem[\protect\citeauthoryear{Wineman and Rajagopal}{Wineman and
  Rajagopal}{2000}]{wineman.as.rajagopal.kr:mechanical}
Wineman, A.~S. and K.~R. Rajagopal (2000).
\newblock {\em Mechanical response of polymers---an introduction}.
\newblock Cambridge: Cambridge University Press.

\end{thebibliography}

\end{document}